\documentclass[journal]{IEEEtran}

\usepackage{amssymb,amsmath,amsthm,enumitem}
        \usepackage{setspace}

\usepackage{graphicx}
\usepackage{caption}
\usepackage{subcaption}
\usepackage[labelformat=parens,labelsep=quad,skip=3pt]{caption}

\ifCLASSINFOpdf
  
\else
 
\fi

\usepackage{amsmath}
\usepackage{amssymb}  
\usepackage{amsfonts}
\usepackage{mathrsfs}
\usepackage{mathptmx} 
\usepackage{mathtools}

\usepackage{xcolor}

\newtheorem{thm}{Theorem}
\newtheorem{lem}{Lemma}

\newtheorem{cor}{Corollary}
\newtheorem{defn}{Definition}

\newtheorem{rem}{Remark}

\newtheorem{assm}{Assumption}

\usepackage{algorithm}
\usepackage{algpseudocode}
\usepackage[algo2e]{algorithm2e} 

\usepackage{multirow}
\usepackage{subcaption}
\usepackage{csquotes}

\begin{document}

\title{Data-driven Safe Control of Uncertain Linear Systems Under Aleatory Uncertainty}

\author{  Hamidreza Modares
\thanks{}
\thanks{
H Modares is with  the Department
of Mechanical Engineering, Michigan State University, East Lansing, MI, 48863, USA, (e-mails: modaresh@msu.edu).}
}

\markboth{}%
{Shell \MakeLowercase{\textit{Modares}}: Data-driven Safe Control of Uncertain Linear Systems Under  Aleatory Uncertainty}

\maketitle

\begin{abstract}
Safe control of constrained uncertain linear systems under aleatory uncertainty is considered. Aleatory uncertainty characterizes random noises and is modeled by a probability distribution function (PDF). Data-based probabilistic safe controllers are designed for the cases where the noise PDF is 1)  zero-mean Gaussian with a known covariance, 2) zero-mean Gaussian with an uncertain covariance, and 3) zero-mean non-Gaussian with an unknown distribution. Easy-to-check model-based conditions for guaranteeing probabilistic safety  are provided for the first case by introducing probabilistic 
$\lambda$-contractive sets. These results are then extended to the second and third cases by leveraging distributionally-robust probabilistic safe control and conditional value-at-risk (\textbf{CVaR}) based probabilistic safe control, respectively.
Data-based implementations of these probabilistic safe controllers are then considered. Moreover, an upper bound on the minimal risk level, under which the existence of a safe controller is guaranteed, is learned using collected data. A simulation example is provided to show the effectiveness of the proposed approach. 

\end{abstract}


\IEEEpeerreviewmaketitle

\section{Introduction} \vspace{-2pt}
\IEEEPARstart{w}{hile} many applications can benefit from increased autonomy (e.g., robots and autonomous vehicles), safety of autonomous systems must be guaranteed before their penetration into society. A challenge in assuring safety is accounting for uncertainties. Two common sources of uncertainty are often referred to as aleatory uncertainty and epistemic uncertainty. The former characterizes the inherent randomness, and the latter characterizes the lack of knowledge. Although aleatory uncertainty cannot be reduced by any additional source of information, epistemic uncertainty can be reduced as more data are collected. In control systems, aleatory uncertainty represents the system and/or measurement noise and is generally modeled by a probability distribution function (PDF). On the other hand,  epistemic uncertainty represents the lack of knowledge on the system dynamics. 

Safe control design methods typically rely on reachability analysis \cite{Reach1}-\cite{Reach7} or control barrier functions (CBFs)  \cite{Inv1}-\cite{Inv7}. To account for aleatory uncertainty, probabilistic CBF-based safe control design has been considered for both continuous-time (CT) \cite{SB1}-\cite{SB6} and discrete-time (DT) systems \cite{SB7}-\cite{SB8}. CBF-based approaches have also been used to certify safety of reinforcement learning algorithms \cite{SafeRL1}-\cite{SafeRL7}. However, due to their computational complexity, the existing CBF-based results for DT systems are limited to finitely supported noise distributions \cite{SB7,SB8}. Moreover, CBF-based methods typically require a complete knowledge of the system dynamics (i.e., they cannot deal with both epistemic and aleatory uncertainties). One way to deal with uncertain dynamics is to identify a model using collected data and leverage it to design model-based safe controllers. However, as shown in this paper, the data requirement conditions for identifying the system dynamics are generally more restrictive than the data requirement conditions for directly learning a safe controller.

Even though safe controllers are designed for uncertain dynamics in \cite{Safedata1,Safedata2}, these approaches ignore the aleatory uncertainty, and, instead, treat it as a bounded disturbance and provide robust safety guarantees. In the presence of aleatory uncertainty, hedging against the worst-case uncertainty to guarantee almost sure (i.e., with probability one) invariance of the safe set may not be feasible, especially when the support of the noise is infinity, large or unknown.



In this paper, data-based probabilistic safe controllers are directly designed for linear discrete-time systems with unknown dynamics affected by noises that are modeled by 1) a zero-mean Gaussian with known covariance, 2) a zero-mean Gaussian with an uncertain covariance, and 3) a zero-mean non-Gaussian with an unknown distribution. 
Probabilistic set invariance guarantees and stability guarantees are unified by introducing probabilistic $\lambda$-contractive sets. It is shown that the probabilistic safety amounts to the value-at-risk (\textbf{VaR}), distributionally robust \textbf{VaR} and conditional value-at-risk (\textbf{CVaR}) based safe control design for cases 1, 2 and 3, respectively. These probabilistic approaches introduce a risk level that specifies how likely a constraint violation may be, i.e., they impose probabilistic or chance constraints instead of deterministic constraints. The risk level depends on the uncertainty level as well as the $\lambda$ parameter. 
Since the minimal risk level is not known a priori, a data-based optimization is provided to learn an upper bound on the minimal risk level. Moreover, data based optimizations are also provided to solve the resulting \textbf{VaR}, distributionally-robust \textbf{VaR} and \textbf{CVaR} based safe control designs.
The contributions of this paper are listed as follows: \vspace{3pt} \\ 
\noindent 1- This paper presents both \textbf{VaR} and \textbf{CVaR} based safe feedback control design methods for stochastic DT systems. These results extend the results of $\lambda$-contractive methods in \cite{Settheoreticcontrol} to stochastic systems with both Gaussian and non-Gaussian noises. \vspace{3pt} \\ 
\noindent 2- In sharp contrast to CBF-based stochastic safe control design methods in \cite{SB7,SB8}, which are limited to finitely-supported noise distributions and must solve a convex optimization at every step, feedback controllers are learned for infinitely-supported noise distributions by solving linear programming optimizations.  
\vspace{3pt} \\
\noindent 3- Data-based optimization problems are developed to not only solve the resulting \textbf{VaR}, distributionally-robust \textbf{VaR} and \textbf{CVaR} based safe control designs, but also to learn the optimal risk level. It is shown that a safe controller can be learned even when the data richness is not enough to identify a system dynamics, and thus CBF-based methods fail.

\vspace{-6pt}
\subsection{Notations and Preliminaries} \vspace{-2pt}
Throughout the paper, $R$ denotes the real numbers. For a matrix $A$, $A_i$ stands for its $i$-th row and $A_{ij}$ is the element of its $i$-th row and $j$-th column. If $A$ and $B$ are matrices (or vectors) of the same dimensions, then $A (\le, \ge) B$ implies a componentwise inequality, i.e., ${A_{ij}} (\le, \ge) {B_{ij}}$ for all $i$ and $j$. 
$Q (\preceq,\succeq)  0$ denotes that  $Q$ is (negative , positive) semi deﬁnite. For a set $\mathcal{A}$, $supp(\mathcal{A})$ stands for its support. $I$ denotes the identity matrix of appropriate dimension. All random variables are assumed to be defined on a probability space $(\Omega,\mathcal{F},Pr)$, with $\Omega$ as the sample space, $\mathcal{F}$ as its associated $\sigma$-algebra and $Pr$ as the probability measure. For a random variable $w: \Omega \longrightarrow R^n$ defined on the probability space $(\Omega,\mathcal{F},Pr)$, with some abuse of notation, the statement $w \in R^q$ is used to state the dimension of the random variable. 
$\mathbb{M}$ denotes the set of all probability measures defined on $(R^p,\mathcal{B})$, with $\mathcal{B}$ the Borel $\sigma$-algebra of $R^q$.  $\mathbb{E}$ indicates the mathematical expectation, and $\mathbb{E}\big[w|z \big]$ denotes the conditional expectation of $w$ with respect to $z$. $w \sim \mathcal{N}(\mu,\Sigma)$ denotes a multivariate Gaussian random vector with the mean $\mu$ and the covariance $\Sigma$. Finally, $\otimes$ represents the Kronecker product. \vspace{6pt}


\begin{defn} \textbf{Polyhedral Set:} \cite{Settheoreticcontrol} polyhedral set 
${\cal {S}} (F,g)$  is represented by \vspace{-9pt}
\begin{align}
{\cal {S}} (F,g) = \{ x \in {R^n}:Fx \le g\} 	,
\label{poly}
\end{align}
where $F \in {R^{q \times n}}$ is a matrix with rows $F_i$, $i=1,...,q$, and $g$ is a vector with elements $g_i$, $i=1,...,q$. 
\end{defn} \vspace{3pt}

\begin{lem} \textbf{Farkas Lemma:} \cite{Farkas} \label{Farkas}
Let $P \in {R}^{m \times n}$, $d \in {R}^{m}$, $c \in {R}^{n}$ and $\beta \in {R}$. Then, the following two statements are equivalent:

 1. There is no $y \in R^m$, such that $P^T y \le c$  and $d^T y > \beta$;

 2. There exists a vector $z \in R^{n}$, such that $z \ge 0$, $Pz=d$, and $c^T z \le \beta$. \hfill   $\square$ 
\end{lem}  \vspace{6pt}



\begin{lem} {\textbf{Probabilistic Lyapunov Stability:}} \cite{Stochstable} \label{stability}
	  	Consider the stochastic system $x(t+1)=h(x(t),w(t))$, where $h(.)$ is a nonlinear function and $w(t)$ is a noise signal. Let $\mathcal{D}$ be a domain containing the origin. Suppose that there exists a continuous function $V : \mathcal{D} \to R$ such that
	\begin{align}
			&V(0)=0, \label{init}\\&
			V(x(t))>0, \,\,\, \forall x(t) \in \mathcal{D}-0 \label{VG}\\&
		 \mathbb{E}[V(x(t+1))]-V(x(t)) \le -c \, V(x(t)), \forall x(t) \in \mathcal{D} \label{Lyap1}
		\end{align}
for some $0 < c < 1$. Then, the origin of the system is exponential stable in probability (ESiP). \hfill   $\square$
\end{lem} \vspace{6pt}





\begin{lem} \label{covdata}
\cite{DRCC} Let $w \sim \mathcal{N}(0,\Sigma)$, where the covariance $\Sigma$ is unknown. Let $L_b=\sup_{w \in supp(\mathcal{N})} ||w||$. Consider the following empirical estimate of the noise covariance $\Sigma$ using $N$ samples
\begin{align} \label{sample1}
\hat \Sigma_N=\frac{1}{N} \sum_{i=1}^N w(i) \, (w(i))^T.
\end{align} 
Then, for the assigned confidence level $\beta \in (0,1)$, if 
\begin{align} \label{Ncond}
N \ge \bigg(2+\sqrt {\left (2 \text{log} \frac{2}{\beta} \right)   }         \bigg)^2,
\end{align}
it holds that
\begin{align} \label{sample2}
||\hat \Sigma_N-\Sigma|| \le  \frac{2L_b^2}{\sqrt{N}}  \bigg ( 2+\sqrt {\left  (2 \text{log} \frac{2}{\beta} \right)  }          \bigg ) =r_c(\beta).
\end{align} \hfill   $\square$
\end{lem} \vspace{-6pt}


\begin{lem} \label{SCC} 
\cite{SCCdeter} Consider a single chance constraint $Pr \big[a^T x+b^T d \le g \big] \ge (1-\epsilon)$, where $x \in R^n$ is the decision variable, $d \sim \mathcal{N}(\mu,\Sigma)$, $a \in R^n$, $b \in R^n$ and $g \in R$. Then, the set of $x$ that satisfies this chance constraint is the convex second-order cone set $\{x:a^T x+b^T \mu \le g-k \sqrt{b^T \Sigma b} \}$ with $k=\sqrt{\frac{1-\epsilon}{\epsilon}}$. \hfill   $\square$
\end{lem}  \vspace{6pt}

\begin{lem} \label{JCC} 
\cite{CCreview}
Consider a joint chance constraint $Pr \big[H x+M d \le g \big] \ge (1-\epsilon)$, where $x \in R^n$ is the decision variable, $d \sim \mathcal{N}(0,\Sigma)$, $H \in R^{q \times n}$, $M \in R^{q \times n}$ and $g \in R^q$.
If the constraints $H_i x+M_i \mu \le g_i-k_i \sqrt{M_i \Sigma  M_i^T}, i=1,...,q$ are satisfied,  where $k_i=\sqrt{\frac{1-\epsilon_i}{\epsilon_i}}$ with $\sum_i \epsilon_i \le \epsilon$, then the original joint chance constraint is also satisfied.
\end{lem}




\section{Problem statement}

Consider a linear discrete-time control system given by
\begin{align}\label{syst}
x(t+1)=Ax(t)+Bu(t)+w(t),
\end{align}
where $x(t) \in R^n$ denotes the system's state at time $t$, $u(t) \in R^m$ denotes the control input, and $w(t) \in R^n$ is an additive random noise. Moreover, $A$ and $B$ are the system matrices of appropriate dimensions and are not known.

\begin{assm} The noise $w$ is either a zero-mean Gaussian noise or a zero-mean non-Gaussian independent and identically distributed (i.i.d) noise.  \end{assm} \vspace{-8pt}

\begin{assm} The pair $(A,B)$ is stabilizable.
\end{assm} \vspace{2pt}



Since the set invariance is used as the key tool for safety guarantee, the following definition is provided. \vspace{6pt}



\begin{defn} \textbf{Positive Invariant Set in Probability (ISiP)}: \cite{Def3} A set $\mathcal{P}$ is an ISiP for the system \eqref{syst} if $x(0) \in \mathcal{P}$ implies that $Pr\big[x(t) \in \mathcal{P}\big]\ge(1-\epsilon) \,\,\, \forall t \ge 0$, where $\epsilon$ is an acceptable risk level. 
\end{defn} \vspace{3pt}

For the case where the noise is Gaussian, the following problem is formulated. \vspace{3pt}

\noindent \textbf{Problem 1} Consider the system \eqref{syst} under Assumptions 1 and 2. Let the noise $w$ be generated by a Gaussian distribution with a known or uncertain covariance. Design a linear feedback controller $u(t)=Kx(t)$ such  that 1) the closed-loop system is ESiP and 2) the polyhedral safe set ${\cal {S}} (F,g)$ is ISiP. \vspace{3pt}

 In the following, we show that Problem 1 imposes chance constraints or \textbf{VaR}-based constraints for safety satisfaction. To this end, note that for the system \eqref{syst} with the control input $u(t)=Kx(t)$, its state at time $t$ can be expressed based on its initial condition $x(0)$ and the noise sequence up to time $t$ as
\begin{align}\label{systresp}
x(t)=(A+BK)^tx(0)+\omega,
\end{align}
where
\begin{align}  \label{omega}
    \omega=(A+BK)^{t-1}w(0)+(A+BK)^{t-2}w(1)+...+w(t).
\end{align}
Therefore, assuring that a polyhedral set is ISiP  is equivalent to imposing a probabilistic constraint in the form of
\begin{align}\label{chance-const}
Pr[F (A+BK)^tx+F \omega \le g] \ge (1-\epsilon), \,\, \forall t \ge 0,
\end{align}
for every $x \in {\cal {S}} (F,g)$. This probabilistic constraint, also referred to as a chance constraint, guarantees that the risk level that the future state trajectories fall outside the polyhedral safe set is at most $\epsilon \in (0,1)$, which is typically near zero.
Chance constraints are closely related to the concept of value-at-risk (\textbf{VaR}) \cite{Var1}, which has been widely used to make risk-aware decisions in many disciplines. To see this, consider the loss function $f(x,\omega)$ where $x \in R^n$ and $\omega \in \Omega$ is a random event ranging over the set $\Omega$ of all random events. 
The \textbf{VaR} of $f(x,\omega)$ at level $\epsilon$ is defined as \cite{Var1}
\begin{align} \label{var}
    \textbf{VaR}_{\epsilon}(x,f)=\inf \bigg\{\lambda \Big| Pr[f(x,\omega) \le \lambda]  \ge \epsilon \bigg\}
\end{align}
Then, one has
\begin{align} \label{var-CC}
    &Pr[f(x,\omega) \le 0] \ge \epsilon \Longleftrightarrow \,\, Pr[f(x,\omega) > 0] \le (1-\epsilon) \\ \nonumber
    &\Longleftrightarrow \,\, \textbf{VaR}_{\epsilon}(x,f) \le 0.
\end{align}

\begin{lem} \label{var-inv}
The set ${\cal {S}} (F,g)$ is ISiP if and only if for every $x \in {\cal {S}} (F,g)$, the following condition is satisfied
\begin{align} \label{var-2}
 \textbf{VaR}_{\epsilon}(x,f) \le 0,
\end{align}
where \vspace{-9pt}
\begin{align} \label{f(x,w)}
f(x,\omega)=F(A+BK)^t x+F\omega ,
\end{align}
and $\omega$ is defined in \eqref{omega}.
\end{lem}
\noindent \textit{Proof.} The proof is immediate from \eqref{var-CC} and \eqref{chance-const}. \vspace{6pt}

However, when the noise distribution is not Gaussian and has fat tails, \textbf{VaR} computations become intractable \cite{CVAR1,Var1}. An alternative risk measure that overcomes these shortcomings is the conditional value-at-risk (\textbf{CVaR}). While \textbf{CVaR} yields the same results in the limited settings where \textbf{VaR} computations are tractable, i.e., for normal distributions, it provides optimization short-cuts which result in tractability even when the noise is non-Gaussian. 

For the loss function, $f(x,\omega)$,  $\textbf{CVaR}_{\epsilon}(x,f)$ is defined as \cite{CVAR1,CVAR2,CVAR3}
\begin{align} \label{cvar1}
    \textbf{CVaR}_{\epsilon}(x,f)=\mathbb{E}\big[z \big| x \ge \textbf{VaR}_{\epsilon}(x,f)  \big],
\end{align}
where $\textbf{VaR}_{\epsilon}(x,f)$ is defined in \eqref{var}. $\textbf{CVaR}_{\epsilon}(x,f)$ can be expressed by a minimization formula \cite{CVAR2}.
\begin{align} \label{cvar2}
    \textbf{CVaR}_{\epsilon}(x,f)=\min_{\eta} F_{\epsilon}(x,\eta),
\end{align}
where \vspace{-10pt}
\begin{align} \label{cvar3}
    F_{\epsilon}(x,\eta)=\eta+\frac{1}{1-\epsilon} \mathbb{E}\big\{ \big[f(x,\omega)-\eta  \big]^{+} \big\},
\end{align}
where $[t]^{+}=\max(0,t)$. \vspace{6pt}

The following problem relaxes the requirement of the noise being Gaussian in Problem 1 and replaces the $\textbf{VaR}_{\epsilon}(x,f)$ constraint of ISiP in \eqref{var-2} with a $\textbf{CVaR}_{\epsilon}(x,f)$ constraint. \vspace{6pt}

\noindent \textbf{Problem 2} Consider the system \eqref{syst} under Assumptions 1 and 2. Let the noise $w$ be generated by an unknown  non-Gaussian PDF. Design a feedback controller $u(t)=K x(t)$ such  that 1) the system is ESiP and 2)   every $x \in {\cal {S}} (F,g)$ satisfies $\textbf{CVaR}_{\epsilon}(x,f) \le 0$ where $f(x,\omega)$ is defined in \eqref{f(x,w)}. \vspace{6pt}

\begin{rem}
Note that by definition of $\textbf{CVaR}_{\epsilon}(x,f)$ in \eqref{cvar1}, one has $\textbf{VaR}_{\epsilon}(x,f) \le \textbf{CVaR}_{\epsilon}(x,f)$. Therefore, if for every $x \in {\cal {S}} (F,g)$ the condition $\textbf{CVaR}_{\epsilon}(x,f) \le 0$ is satisfied, then $\textbf{VaR}_{\epsilon}(x,f) \le 0$ is also satisfied and thus based on Lemma \ref{var-inv}, the set ${\cal {S}} (F,g)$ is ISiP. 
\end{rem}

\section{Probabilistic contractive sets for probabilistic stability and safety} 
To solve Problems 1 and 2, the following definition of probabilistic contractive sets is introduced, as an extension of its deterministic counterpart in \cite{Settheoreticcontrol}. \vspace{3pt}

\begin{defn} \textbf{Contractive Sets in Probability:} Given a $\lambda \in (0,1)$, the set  $\cal{P}$ is $\lambda$-contractive in probability for the system \eqref{syst} if $x(t) \in \cal{P}$ implies that  $Pr[x(t+1) \in \lambda \cal{P}] \ge$ $(1-\epsilon)$  where $\epsilon$ is a risk level. \vspace{6pt}
\end{defn}

The following lemmas make the connection between probabilistic contractive sets and ISiP. Before proceeding, for a random variable $w$, define the following optimization problem 
\begin{align} 
& \bar h= \text{Opt}(H_w,\epsilon)= \min \sum_{j} {{ h}_j} \nonumber \\ 
& s.t. \,\, Pr[H_w \,\, w(t) \le { h}]=(1-\epsilon). \label{boundcon}
\end{align}
where $H_w$ is a matrix and $h$ is a vector. Then, for any matrix $L$ and decision variable $x$,
\begin{align} \label{cce1}
    Pr\big[L x(t)+ H_w w(t)\big)
\le g \big] \ge (1-\epsilon),
\end{align}
is equivalent to \cite{opt11}
\begin{align}  \label{cce2}
    L x(t) \le  g - \bar h,  \,\, \bar h=\text{Opt} (H_w,\epsilon).
\end{align} 
Note that as $\epsilon$ increases, $\bar h$ decreases. 
\vspace{6pt}

\begin{lem} \label{lem1}
Consider the system \eqref{syst} with $u(t)=Kx(t)$. If a polyhedral set  ${\cal{S}}(F,g)$ is a  $\lambda$-contractive set in probability with a risk level $\epsilon_1$, then $x(t) \in {\cal{S}}(F,g)$ implies that $Pr[x(t+1) \in {\cal{S}}(F,g)] \ge$ $(1-\epsilon_2)$ for some  $\epsilon_2 < \epsilon_1$. 
\end{lem}  \vspace{2pt}
\textit{Proof.} Let $x(t) \in {\cal{S}}(F,g)$. Since ${\cal{S}}(F,g)$ is a $\lambda$-contractive set in probability with the risk level $\epsilon_1$, by definition it implies that  $Pr\big[F \big( (A+BK)x(t)+ w(t)\big)
\le \lambda g \big] \ge$ $(1-\epsilon_1)$. Based on the equivalence of \eqref{cce1} and \eqref{cce2}, this is equivalent to $F  (A+BK)x(t) 
\le \lambda g - \bar h_1=h_1$ where $\bar h_1=\text{Opt} (F,\epsilon_1)$. On the other hand, since $\lambda \in (0,1)$, $\lambda g - \bar h_1=g - \bar h_2=h_1$ for some $\bar h_2 > \bar h_1$. Therefore, based on the equivalence of \eqref{cce1} and \eqref{cce2}, $F ( (A+BK)x(t) 
\le g - \bar h_2= h_1$ implies that $Pr\big[F \big( (A+BK)x(t)+ w(t)\big)
\le g \big] \ge$ $(1-\epsilon_2)$ for some $\epsilon_2$ satisfying $\bar h_2=\text{Opt}(F,\epsilon_2)$, which guarantees $Pr[x(t+1) \in {\cal{S}}(F,g)] \ge$ $(1-\epsilon_2)$. Moreover, based on \eqref{boundcon}, since $\bar h_2 > \bar h_1$, it implies that  $\epsilon_2 < \epsilon_1$. \hfill   $\square$ \vspace{6 pt}

\begin{lem} \label{cont-inv-eq0}
If the set ${\cal{S}}(F,g)$ is a $\lambda$-contractive set in probability, then it is also ISiP.
\end{lem} \vspace{2pt}
\textit{Proof.} Let the set ${\cal{S}}(F,g)$ be a $\lambda$-contractive set in probability with the risk level $\epsilon_1$. Then, based on Lemma \ref{lem1}, $Pr \big[x(t) \in {\cal{S}} (F,g)\ \big| \, x(t-1) \in {\cal S}(F,g) \big] \ge (1-\epsilon_2)$ for some  $\epsilon_2 < \epsilon_1$. Using this property, and since the noise is i.i.d by Assumption 1, and thus $x(t)$ has a Markov property, one has
\begin{align*}
&Pr \big[x(2) \in {\cal{S}} (F,g) \big|  x(0) \in {\cal S} (F,g) ,x(1) \in {\cal S}(F,g) \big] \\ \nonumber
&=Pr \big[x(2) \in {\cal{S}}(F,g) \big| x(1) \in {\cal S}(F,g) \big] \\ \nonumber
&\times Pr \big[x(1) \in {\cal{S}}(F,g)  \big| x(0) \in {\cal S}(F,g) \big] \ge (1-\epsilon_2)^2. 
\end{align*}

Using this recursive reasoning, one has $Pr \big[x(t) \in {\cal{S}}(F,g) \,\ \big| \, x(0) \in {\cal S}(F,g) \big] \ge (1-\epsilon_2)^t=(1-\epsilon_3)$ for some $\epsilon_3$, which can be greater than or less than $\epsilon_1$, depending on $\lambda$ and $t$. \hfill   $\square$ \vspace{6pt}

\begin{rem}
Note that for large values of $t$, the risk level can become large and unacceptable. However, since the planning horizon for any control system is generally finite, $\lambda$-contractivity guarantees that a set is ISiP with an acceptable risk level for a time duration longer than the planning horizon. 
\end{rem} \vspace{6pt}



The next results provide conditions under which the probabilistic contractivity of  ${\cal S}(F,g)$ is guaranteed.

To provide conditions for probabilistic $\lambda$-contractiveness, select $\epsilon_i, i=1,...,q$ such that $\sum_{i=1}^q \epsilon_i \le \epsilon$ and define $l=[l_1,...,l_q]$ where
\begin{align} \label{l1}
l_i=\sqrt{\frac{1-\epsilon_i}{\epsilon_i}}  \sqrt{F_i  \Sigma F_i^T}.
\end{align}

We now define the following operator for the system \eqref{syst} with the safe set $ {\cal S} (F,g)$ under the state-feedback control $u(t)=Kx(t)$.
\begin{align} \label{before2}
{\cal S}_{Pr}^-(F,g)=\{x | F(A+BK)x \le \lambda g-l \}.
\end{align}
which is the set of all previous states for which it is guaranteed that their current states lie inside $\lambda {\cal S}(F,g)$ with a probability of at least $1-\epsilon$.

\begin{lem} \label{JCC2} 
Consider the system \eqref{syst} under Assumptions 1 and 2 with $w=\mathcal{N}(0,\Sigma)$. Let the control input  be $u(t)=K x(t)$. Then, the polyhedral set ${\cal {S}} (F,g)$ is $\lambda$-contractive in probability with the risk level $\epsilon$ if \vspace{-6pt}
\begin{align} \label{expect-eq2}
    {\cal S} (F,g)\subseteq {\cal S}_{Pr}^-(F,g).
\end{align}
\end{lem} \vspace{3pt}
\textit{Proof.} By Lemma \ref{JCC}, if $F(A+BK)x \le \lambda g-l$, the joint chance constraint $Pr[F(A+BK)x+Fw \le \lambda g] \ge (1-\epsilon) $ is satisfied. Therefore, the set \eqref{before2} is a safe underestimation of the set of all previous states for which it is guaranteed that their current state lies inside $\lambda {\cal S}$ with a probability of at least $1-\epsilon$. Therefore,   $x(t) \in {\cal S}_{Pr}^-(F,g)$  implies that $Pr[x(t+1) \in \lambda {\cal S}(F,g)] \ge (1-\epsilon)$, which proves that \eqref{expect-eq2} is a sufficient condition for $\lambda$-contractive in probability. \hfill   $\square$ \vspace{6pt}

\begin{rem} \label{CC-comp}
Based on Lemma \ref{SCC}, the sufficient condition \eqref{expect-eq2} becomes a necessary and sufficient condition for single chance constraints, i.e., when $g$ is a scalar. The joint chance constraint is split into multiple single chance constraints in Lemma \ref{JCC} at the price of conservativeness introduced in the inequality, and therefore, the condition \eqref{expect-eq2} only provides a sufficient condition. 
\end{rem} \vspace{6pt}

The previous results assumed that the covariance of the noise is known. However, the noise covariance is generally unknown and must be approximated using data samples. In this case, only a certain number $N$ of independent realizations of the random vector $w$ are available which are used to find the empirical estimate of the covariance. The covariance estimate, however, cannot be accurately found using a finite number of samples, and, instead, belongs to an ambiguity set. Based on Lemma \ref{covdata}, for an arbitrarily-chosen confidence level $\beta \in (0,1)$, its corresponding ambiguity set $\mathbb{A}$ is defined as
\begin{align} \label{ambig}
    \mathbb{A}:=\bigg\{Pr \in \mathbb{M} \bigg| \mathbb{E}\big[(w)(w)^T\big] \le r_b(\beta) \hat \Sigma_N                     \bigg\},
\end{align} 
where $\hat \Sigma_N$ and $r_b(\beta)$  are obtained based on \eqref{sample1} and \eqref{sample2}, respectively, in Lemma \ref{covdata}. The controller is then designed such that 
\begin{align}
    \inf_{Pr \in \mathbb{A}}\Big[Pr \big[x(t) \in \mathcal{S} \big] \ge (1-\epsilon) \Big] \ge (1-\beta), \,\, \forall t \ge 0.
\end{align}
These constraints are referred to as distributionally robust chance constraints, as the constraints must hold with a given confidence level for all disturbance distributions that belong to the ambiguity set. \vspace{6pt}

Select $\epsilon_i, i=1,...,q$ such that $\sum_{i=1}^q \epsilon_i \le \epsilon$ and define $\hat l=[\hat l_1,...,\hat l_q]$ where
\begin{align} \label{l2}
\hat l_i=\sqrt{\frac{1-\epsilon_i}{\epsilon_i}}  \sqrt{F_i (\hat \Sigma_N+r_c(\beta)) F_i^T}.
\end{align}
Now, similar to \eqref{before2}, define 
\begin{align} \label{before3}
\hat {\cal S}_{Pr}^-(F,g)=\{x | F(A+BK)x \le \lambda g-\hat l \}.
\end{align}

\begin{lem} \label{JCC3} 
Consider the system \eqref{syst} under Assumptions 1 and 2 with $w \sim \mathcal{N}(0,\Sigma)$ where $\Sigma$ is unknown and its sample average $\hat \Sigma_N$ is calculated using \eqref{sample1}. Let the control input  be $u(t)=K x(t)$. Fix  $\beta \in (0,1)$. 
Then, with a probability of at least $1-\beta$  the polyhedral set ${\cal {S}} (F,g)$ is $\lambda$-contractive in probability if 
\begin{align} \label{cont2}
        {\cal S} (F,g)\subseteq \hat {\cal S}_{Pr}^-(F,g).
\end{align}
\end{lem} \vspace{3pt}
\textit{Proof.} The proof is based on Lemma \ref{JCC} and is similar to the proof of Lemma \ref{JCC2}. \hfill   $\square$ \vspace{4pt}

The following theorem and corollaries show that Problems 1 and 2 can be solved by making the polyhedral set ${\cal {S}} (F,g)$ a probabilistic $\lambda$-contractive set for the closed-loop system.  \vspace{5pt}

\begin{thm} \label{stability0}
Consider the system \eqref{syst} under Assumptions 1 and 2. Let $w \sim \mathcal{N}(0,\Sigma)$. Then, a controller $u(t)=K x(t)$ that makes the set ${\cal {S}} (F,g)$ $\lambda$-contractive in probability 
by satisfying \eqref{expect-eq2}
guarantees that the system is ESiP and the set ${\cal {S}} (F,g)$ is ISiP. Therefore, it solves Problem 1. 
\end{thm}
\noindent \textit{Proof.} Let $V:{\cal {S}} (F,g) \to$ $\mathbb{R}$ be defined as
\begin{align}\label{lyap}
V(x(t)):=\max_{i \in {1,..,q}} \Big|\frac{F_i x(t)}{g_i} \Big|,
\end{align}
for which we have $V(x)>0$ for all $x \in \cal {S}$ and $V(x)=0$ if and only if $x=0$. We first show that if the set ${\cal {S}} (F,g)$ is a probabilistic $\lambda$-contractive set for the closed-loop system, then \eqref{lyap} is a Lyapunov function for the closed-loop system satisfying
\begin{align}\label{lyap2}
\mathbb{E}V[x(t+1)]- V(x(t)) \le -(1-\lambda) V(x(t)),
\end{align}
where $(1-\lambda) > 0$ since $\lambda \in [0,1)$, and, therefore, based on Lemma \ref{stability}, it guarantees ESiP. For an arbitrary $x(t) \in {\cal {S}} (F,g)$, $F_i x(t) \le g_i, i=1,...,q$, or equivalently $\Big|\frac{F_ix(t)}{g_i} \Big| \le 1, i=1,...,q$. Therefore, based on \eqref{lyap}, one has $V(x(t))=c$ for some $c \le 1$. This implies that $Fx(t) \le cg$ or equivalently $x(t) \in {\cal {S}} (F,cg)$. Moreover, if ${\cal {S}} (F,g)$ is $\lambda$-contractive in probability, then  ${\cal {S}} (F,cg)$ is also $\lambda$-contractive in probability.
Based on \eqref{expect-eq2} and \eqref{before2}, the $\lambda$-contractiveness in probability of ${\cal {S}} (F,cg)$ implies that if $x(t) \in {\cal {S}} (F,cg)$, then $F(A+BK)  x(t)=\mathbb{E}V(x(t+1)) \le c \lambda- \, l = \lambda V(x(t))-  \, l \le \lambda V(x(t))$, where $l=[l_1,...,l_q]$ with $l_i$ being defined in \eqref{l1}. Using \eqref{lyap}, this is equivalent to \eqref{lyap2}. This proves stability for $\lambda$-contractive in probability. The probabilistic safety guarantee for $\lambda$-contractive in probability is shown in Lemma \ref{cont-inv-eq0}. This completes the proof. \hfill   $\square$ \vspace{6pt}

\begin{cor} \label{stability2}
Consider the system \eqref{syst} under Assumptions 1 and 2 with $w \sim \mathcal{N}(0,\Sigma)$ where $\Sigma$ belongs to the ambiguity set \eqref{ambig} with a probability of at least $1-\beta$. Then, a controller $u=Kx$ that makes the set ${\cal {S}} (F,g)$ $\lambda$-contractive in probability guarantees that the system is ESiP and that the safe set ${\cal {S}} (F,g)$ is ISiP with a probability of at least $1-\beta$.  Therefore, it solves Problem 1 with a probability of at least $1-\beta$.
\end{cor}  \vspace{2pt}
\noindent \textit{Proof.} The proof uses \eqref{cont2} and follows the same procedure as the proof of Theorem \ref{stability0}.  \hfill   $\square$ \vspace{6pt}

\begin{cor} \label{stability3}
Consider the system \eqref{syst} under Assumptions 1 and 2 with $w$ as a non-Gaussian noise with an unknown PDF. Then, a controller $u(t)=K x(t)$ that satisfies the \textbf{CVaR} condition
\begin{align} \label{cvarinv}
x \in {\cal {S}} (F,g) \implies \textbf{CVaR}_{\epsilon}(x,f_0) \le 0 , 
\end{align}
where 
\begin{align} \label{f0}
 f_0(x,w)=F (A+BK)x+Fw- \lambda g,
\end{align}
guarantees that the system is ESiP and that the safe set ${\cal {S}} (F,g)$ is ISiP.  Therefore, it solves Problem 2.
\end{cor}  \vspace{2pt}
\noindent \textit{Proof.} Based on $\textbf{VaR}_{\epsilon}(x,f) \le \textbf{CVaR}_{\epsilon}(x,f)$, the condition \eqref{cvarinv} assures that  the condition  $\textbf{VaR}_{\epsilon}(x,f_0) \le 0$ is satisfied, which is equivalent to $Pr[F x(t+1) \le \lambda g] \ge (1-\epsilon)$. Therefore, the condition \eqref{cvarinv} guarantees that the set ${\cal {S}} (F,g)$ is $\lambda$-contractive in probability. Based on Theorem \ref{stability0}, the proof is completed.  \hfill   $\square$ \vspace{5pt}




\section{Conditions for model-based probabilistic safety guarantees} 
In this section, easy-to-check conditions under which a set can be made $\lambda$-contractive in probability (Theorems \ref{modelcondVar}  and \ref{distrobust} for Gaussian distributions with known and uncertain covariances, respectively, and Theorem \ref{modelcondCVar} for non-Gaussian distributions) are provided. Moreover, an upper bound on the minimum achievable risk level is found. \vspace{8pt}

\begin{thm} \label{modelcondVar}
\textbf{Gaussian noise with known covariance:} Consider the system \eqref{syst} under Assumptions 1 and 2 with $w \sim \mathcal{N}(0,\Sigma)$.   Let $u(t) = K x(t)$. Then, the polyhedral set ${\cal{S}}(F,g)$   is  $\lambda$-contractive in probability if there exists a nonnegative matrix 
$P$ such that
\begin{align} \label{conteq11}
&{P}F = F(A + BK)\\ \nonumber
&{\rm{ }}{P}g  \le \lambda g-l.
\end{align} 
\end{thm} \vspace{3pt}
where $l=[l_1,...,l_q]$ with $l_i$ being defined as \eqref{l1}. \vspace{3pt}

\noindent \textit{Proof.} Based on Lemma \ref{JCC2},  ${\cal{S}}(F,g)$ is $\lambda$-contractive in probability if ${\cal{S}}(F,g) \subseteq \{x:F (A+BK) x \le \lambda g-l \big \}$. This, in turns, implies that $\{x:F x \le g\} \cap \{x:\big(F (A+BK) \big)_i \, x > \lambda g_i- l_i \}=\emptyset, \,\, \forall i=1,...,q $. Based on the Farkas lemma in Lemma \ref{Farkas}, this is equivalent to the existence of a vector $p_i \ge 0$ such that $F^T p_i=\big(F (A+BK) \big)_i$ and $g^T p_i \le \lambda g_i-l_i$. Defining $P=[p_1,..,p_q]^T$, one gets \eqref{conteq11}. \hfill   $\square$ \vspace{6pt}


\begin{thm} \label{distrobust}
\textbf{Gaussian noise with uncertain covariance:}  Consider the system \eqref{syst} under Assumptions 1 and 2 with $w \sim \mathcal{N}(0,\Sigma)$ where $\Sigma$ is unknown. Let $u(t) = K x(t)$. Let $N$ i.i.d samples of the zero-mean noise $w$ be collected and its empirical covarianace be calculated using \eqref{sample1}. Let $N$ satisfy the condition \eqref{Ncond} and thus with a probability of at least $(1-\beta)$, the covariance belongs to the ambiguity set \eqref{ambig}. Then, with a confidence level of $(1-\beta)$, the ${\cal {S}} (F,g)$ is a $\lambda$-contractive set in probability  with the risk level $\epsilon$ if there  exists a nonnegative matrix 
$P$ such that \vspace{-6pt}
\begin{align} \label{conteq-rob}
&{P}F = F(A + BK)\\ \nonumber
&{\rm{ }}{P}g  \le \lambda g-\hat l.
\end{align} 
where $\hat l=[\hat l_1,...\hat l_q]$ with $\hat l_i$ being defined as \eqref{l2}. 
\end{thm} \vspace{3pt}
\noindent \textit{Proof.} The proof is similar to Theorem \ref{modelcondVar} and based on Lemma \ref{JCC3} instead of Lemma \ref{JCC2}.  \hfill   $\square$ 



Theorems  \ref{modelcondVar} and \ref{distrobust} require to assign a risk level a priori. The risk level $\epsilon$, however, depends on the uncertainty level as well as the $\lambda$ parameter. 
Since the optimal (minimal) risk level is not known a priori, the following theorem provides an upper bound on the minimal risk level based on the steady-state covariance of the system's state for the case under which the noise covariance is unknown and belongs to an ambiguity set.  

\vspace{6pt}

\begin{thm} \label{risklevel}
\textbf{Risk bound:} Consider the system \eqref{syst} under Assumptions 1 and 2 and with a control input $u(t)=Kx(t)$. Let $N$ samples of a Gaussian noise be collected and its empirical covarinace be calculated using \eqref{sample1}. Let $N$ satisfy the condition \eqref{Ncond} and thus with a probability of at least $(1-\beta)$, the covariance belongs to the ambiguity set \eqref{ambig}. Then, with a probability of at least $(1-\beta)$, the solution $\bar \epsilon$ to the following  optimization problem represents a bound on the lowest risk level for guaranteeing that the set ${\cal {S}} (F,g)$ is $\lambda$-contractive in probability.     \vspace{-6pt}                               \begin{subequations}
\begin{align} \label{riska}
& \min_{V,\Sigma_{ss}^b, \epsilon}  \epsilon \\ 
& \text{s.t.} \,\,\, 
\begin{bmatrix}
\Sigma_{ss}^b- \big(\hat \Sigma_{N}+r_c(\beta)\big)  & (A \Sigma_{ss}^b+BV) \label{riskb} \\
(A \Sigma_{ss}^b+BV)^T & \Sigma_{ss}^b
\end{bmatrix} \succeq 0 \\ 
& \frac{q}{6\lambda^2 g_i^2} F_i \Sigma_{ss}^b F_i^T \le  \epsilon, i=1,..,q \label{epsN} \\ 
& {\Sigma_{ss}^b} \succeq 0,
\end{align} 
 \end{subequations}
Moreover, this risk level is achieved by the controller gain $K=V {\Sigma_{ss}^b}^{-1}$. 
\end{thm} \vspace{2pt}

\noindent \textit{Proof.} Since $A+BK$ is strictly stable, the state trajectories of the closed-loop system $x(t+1)=(A+BK)x(t)+w(t)$ converge to a stationary distribution for which its covariance $ \bar \Sigma_{ss}^x=\mathbb{E} \Big[ x x^T \Big]$ satisfies \cite{Morari} \vspace{-6pt}
\begin{align}
    \bar \Sigma_{ss}^x=(A+BK) \bar \Sigma_{ss}^x (A+BK)^T+\Sigma,
\end{align}
where $\Sigma$ is the actual covariance of the noise $w$. Since $\Sigma$ is not known and it is only known that with a probability of $1-\beta$ it belongs to the ambiguity set \eqref{ambig}, then $\Sigma \preceq (\hat \Sigma_N+r_c(\beta))$ with a probability of at least $1-\beta$. Therefore, the solution to the following Lyapunov equation
\begin{align}
    \bar \Sigma_{ss}^b = (A+BK) \bar \Sigma_{ss}^b (A+BK)^T+(\hat \Sigma_N+r_c(\beta)), 
\end{align}
satisfies $\bar \Sigma_{ss}^b \succeq \bar \Sigma_{ss}^x$ with a probability of at least $1-\beta$. On the other hand,  the solution to the inequality 
\begin{align} \label{lyapineq1}
    \Sigma_{ss}^b  \succeq (A+BK)\Sigma_{ss}^b (A+BK)^T+ (\hat \Sigma_N+r_c(\beta)) 
\end{align}
is an upper bound of $\bar \Sigma_{ss}^b$. By defining $K=V {\Sigma_{ss}^b}^{-1}$ and using Schur complement, \eqref{lyapineq1} is equivalent to the linear matrix inequality (LMI)
\begin{align}
\begin{bmatrix}
\Sigma_{ss}^b- \big(\hat \Sigma_{N}+r_c(\beta)\big)  & (A \Sigma_{ss}^b+BV)\\
(A \Sigma_{ss}^b+BV)^T & {\Sigma_{ss}^b}
\end{bmatrix} \succeq 0
\end{align}
Minimizing over the feasible space of this LMI solutions, i.e, the best upper bound of $ \Sigma_{ss}^b$, is equal to $ \bar \Sigma_{ss}^b$, and, therefore, any solution to the optimization problem is also an upper bound to $\bar \Sigma_{ss}^x$ with a probability of at least $1-\beta$. On the other hand, based on Lemma \ref{JCC}, $Pr[Fx \le g] \ge (1-\epsilon)$ is satisfied if $Pr[F_i x \le g_i] \ge (1-\bar \epsilon_i)$ and $\sum_{i=1}^q \bar \epsilon_i \le \epsilon$. Set $\epsilon_i=\frac{\epsilon}{q}$. Using Chebyshev's inequality \cite{VCDdim}, and when the state reaches the stationary condition, one has $Pr[F_i x > g_i] \le \frac{F \Sigma_{ss}^x F^T}{6\lambda^2 g_i^2}$. Therefore, if $\frac{1}{6\lambda^2 g_i^2} F_i \Sigma_{ss}^b F_i^T \le \frac {\epsilon}{q}, i=1,..,q$, then the original chance constraint is satisfied in the steady state. This completes the proof. \hfill   $\square$  \vspace{3pt}

\begin{rem}
For the case where the covariance $\Sigma$ is known, the worst-case covariance $\Sigma_{N}+r_c(\beta)$ is replaced with the actual covariance $\Sigma$ in Theorem \ref{risklevel} and the confidence probability $1-\beta$ becomes 1. Moreover, for the single-chance constraint case, i.e., when $g$ is scalar, \eqref{epsN} reduces to only one equation. 
\end{rem} \vspace{3pt}

When the noise distribution is completely unknown, Problem 2, for which a \textbf{CVaR} constraint is imposed, must be solved. Since the exact evaluation of the expectation in \textbf{CVaR} is difficult due to the piecewise linearity
of the operator $[.]^{+}$, it is typically approximated using sample average approximation methods based on $N$ available i.i.d scenario data of the noise \cite{datacvar} and thus the \textbf{CVaR} condition \eqref{cvarinv} is approximated by \vspace{-6pt}
\begin{align} \label{cvar4}
   &\widehat{CVaR_{\epsilon}}^N=\min_{\eta_i,z_i} \,\, \eta+\frac{1}{1-\epsilon} \frac{1}{N} \sum_{i=1}^N  z_i  \\ \nonumber
   & \text{s.t.} \,\, z_i \ge f_0(x,w_i)-\eta, \,\, z_i \ge 0, \,\, i=1,..,N,
\end{align}
where $f_0$ is defined in \eqref{f0} and $\widehat{CVaR_{\epsilon}}^N$ is the empirical \textbf{CVaR}. \vspace{4pt}

\begin{thm} \label{modelcondCVar}
\textbf{non-Gaussian noise:} Consider the system \eqref{syst} under Assumptions 1 and 2 and let $N$ i.i.d samples of the noise $w_i, \,\, i=1,...,N$ be available. Let $u(t) = Kx(t)$. Then, the polyhedral set ${\cal{S}}(F,g)$ with $g \in R$ is  $\lambda$-contractive in probability with a confidence level depending on $N$ if and only if there exists a nonnegative matrix 
$P$ such that
\begin{align} \label{conteq15}
&\min_{\eta,z_i} \bigg( \eta+\frac{1}{N(1-\epsilon)} \sum_{i=1}^N z_i \bigg)  \le 0 \\ \nonumber
&{P}F = F(A + BK)\\ \nonumber
&{\rm{ }}{P}g  \le \lambda g-F w_i+z_i+\eta, \,\, i=1,...,N \\ \nonumber
& z_i \ge 0.
\end{align} 
\end{thm}

\noindent \textit{Proof.} Based on Corollary \ref{stability3} and the approximation \eqref{cvar4},  ${\cal{S}}(F,g)$ is $\lambda$-contractive in probability if ${\cal{S}}(F,g) \subseteq \{x:F (A+BK) x+Fw_i \le \lambda g+z_i+\eta\}$ where $\eta$ and $z_i$ satisfy \eqref{conteq15}. Using Farkas lemma completes the proof. \hfill   $\square$ \vspace{6pt}

\begin{rem}
While Theorem \ref{modelcondCVar} is presented for a single constraint, the joint \textbf{CVaR} constraints can also be handled similar to the joint chance constraints in  Lemma \ref{JCC2} by splitting joint constraints into single constraints. Moreover, probably approximately correct (PAC) \cite{VCDdim} data-based confidence levels can be found using the sample average estimation of \textbf{CVaR} \cite{CVaRsample}. 
\end{rem}
\section{Data-based safe risk assessment and control design}

We assume that a data set of $N$ input/state and noise measurements are collected from the system \eqref{syst} by applying a sequence
${u}(0),...,{u}(N- 1)$ of input and measuring the corresponding state values 
${X} = \left[ {{x}(0),...,{x}(N)} \right]$ and the noise values ${w}(0),...,{w}(N - 1)$. Let these data samples be arranged as
\begin{subequations}
\begin{align} \label{datacollected}
  &{U_0} = \left[ {{u}(0),...,{u}(N - 1)} \right] \\
  &{W_0} = \left[ {{w}(0),...,{w}(N- 1)} \right]  \\
  &{X_0} = \left[ {{x}(0),...,{x}(N - 1)} \right]  \\
  & {X_1} = \left[ {{x}(1),...,{x}(N)} \right]
   \label{datacollected2}
   \end{align}
\end{subequations}

As shown in the next theorem, and supported in the simulation results, learning a data-based safe controller directly is less data intensive than learning a system model that CBF-based methods rely on. Therefore, in low data regimes, the presented approach is more advantageous to CBF-based methods.


\begin{thm} \textbf{Data-based versions of Theorems \ref{modelcondVar} and \ref{distrobust}:} \label{datareq}
Consider the system \eqref{syst} under Assumptions 1 and 2 with $w \sim \mathcal{N}(0,\Sigma)$. Let the input/output/noise data be collected from applying an open-loop control sequence to the system and arranged by \eqref{datacollected}-\eqref{datacollected2}. Let the data matrix $X_0$ be full row rank. Then, there exists a controller $u(t)=K x(t)$ that makes the set ${\cal{S}}(F,g)$ a probabilistic $\lambda$-contractive set if there exist matrices
 $G_k$  and $P$ satisfying \vspace{-3pt}
\begin{align}  \label{data2}
&P g  \le \lambda g-a,\\ \nonumber
&P F = F(X_1-W_0) G_k. \\ \nonumber
& X_0 \, G_K=I
\end{align}
where $a=l$ with $l$ being defined in \eqref{l1} when the noise covariance is known, and $a=\hat l$ with $\hat l$ being defined in \eqref{l2} when the noise covariance is estimated through samples.  Moreover, the control gain that solves Problem 1 is $K=U_0 \, G_K$.
\end{thm}

\vspace{2pt}

\noindent \textit{Proof.} Since the matrix $X_0$ is assumed full rank, a right inverse $G_{K}$ exists such that $X_0 \, G_{K}=I$. 
Based on the data collected in \eqref{datacollected}-\eqref{datacollected2} and the stochastic linear system \eqref{syst}, one has
\begin{align} \label{data lpv}
{X_1-W_0} = {A}{X_0} + B{U_0}.
\end{align}
Multiplying both sides of \eqref{data lpv} by $G_{K}$ from right yields
\begin{align}\label{data form}
({X_1-W_0})G_{K} = {A} + B{U_0}G_{K}.
\end{align}
Using the control gain ${K} = {U_0}G_{K}$, one obtains $A+BK=(X_1-W_0) \, G_K$
. Therefore, ${P}F = F(A + BK)$ becomes $P F = F(X_1-W_0) G_k$. The rest of the proof follows Theorems \ref{modelcondVar} and \ref{distrobust} for the first case and second case, respectively. \hfill   $\square$ \vspace{-9pt}

\begin{rem}
The optimization problem \eqref{data2} is a linear programming (LP), which can be efficiently solved. The value of $\lambda$ can be either pre-determined or minimized over in the LP optimization.  While it is desired to find the optimal boundary-crossing speed for some applications, it is also desired to fix it to a large value close to one if the safe control is to be merged with a nominal controller to minimize the intervention with the nominal controller. 
\end{rem} \vspace{3pt}

\begin{cor} \label{datareq2}
\textbf{Data-based version of Theorem \ref{modelcondCVar}:} \label{datareq}
Consider the system \eqref{syst} under Assumptions 1 and 2. Let the noise distribution $w$ be unknown. Let the input/output/noise data be collected from applying an open-loop control sequence to the system and arranged by \eqref{datacollected}-\eqref{datacollected2}. Let the data matrix $X_0$ be full row rank. Then, there exists a controller $u(t)=Kx(t)$ to make the set ${\cal{S}}(F,g)$ a probabilistic $\lambda$-contractive set if there exist matrices
 $G_k$  and $P$ satisfying
%
\begin{align} \label{data3}
& \min_{\eta,z_i} \bigg( \eta+\frac{1}{N(1-\epsilon)} \sum_{i=1}^N z_i \bigg)  \le 0 \\ \nonumber
& {P}g  \le \lambda g-{F w_i}+z_i+\eta, \,\,  i=1,...,N \\ \nonumber
& z_i \ge 0 \\ \nonumber
& P F = F(X_1-W_0) G_k \\ \nonumber
& X_0 \, G_K=I
\end{align}
Moreover, the control gain that solves Problem 2 is $K=U_0 \, G_K$.

\noindent \textit{Proof.} The proof is similar to the proof of Theorem 6 and uses the results of Theorem \ref{modelcondCVar}. \hfill   $\square$ \end{cor}
\vspace{6pt}  

\begin{rem} 
Even though the data informativeness requirement for learning a safe control is expressed in \cite{Safedata2} as that the matrix 
\begin{align} \label{mat} \left[ \begin{array}{l}
{U_0}\\
{X_0}
\end{array} \right] 
\end{align}
has full row rank, it was shown in Theorem 6 that a weaker data-informativeness condition is required. If the matrix \eqref{mat} is full rank (i.e, the data is persistently exciting (PE)), then the  deterministic system \eqref{syst} with $w=0$ can be uniquely identified. However, based on Theorem 6, the PE requirement is not needed to design a safe controller. Therefore, in many situations, it is desirable to directly design a safe controller using data rather than first attempting to identify the system and then designing a controller based on the identified models. 
\end{rem} \vspace{6pt}

The data-based version of Theorem \ref{risklevel} is presented next to learn an upper bound on the optimum risk level using only measured data. \vspace{3pt}

\begin{thm} \label{riskaa}
 Consider the system \eqref{syst} and let conditions of Theorem \ref{risklevel} be satisfied. Let the input/output/noise data be collected from applying an open-loop control sequence to the system and arranged by \eqref{datacollected}-\eqref{datacollected2}. Then, with a probability of at least $(1-\beta)$, the solution $\bar \epsilon$ to the following data-based optimization problem represents a bound on the lowest risk level for guaranteeing $\lambda$-contractive in probability of the set ${\cal {S}} (F,g)$, 
  \begin{subequations}
\begin{align} \label{riskda}
& \bar \epsilon= \min_{V,\Sigma_{ss}^b, \epsilon}  \epsilon \\ 
& \text{s.t.} \,\,\, 
\begin{bmatrix}
\Sigma_{ss}^b- \big(\hat \Sigma_{N}+r_c(\beta)\big)  & (X_1-W_0)V)\\
( (X_1-W_0)V))^T & \Sigma_{ss}^b
\end{bmatrix} \succeq 0 \label{riskda2} \\ 
& \frac{q}{6\lambda^2 g_i^2} F_i \Sigma_{ss}^b{F_i^T} \le \epsilon, i=1,..,q \\ 
& \Sigma_{ss}^b \succeq 0. 
\end{align} 
 \end{subequations}
which is achieved by the controller gain $K=U_0 V {\Sigma_{ss}^b}^{-1}$. \hfill   $\square$
\end{thm} \vspace{2pt}
\noindent \textit{Proof.} The condition \eqref{riskb} is equivalent to
\begin{align}
\begin{bmatrix}
\Sigma_{ss}^b- \big(\hat \Sigma_{N}+r_c(\beta)\big)  & (A+BK) \Sigma_{ss}^b)\\
((A+BK)\Sigma_{ss}^b)^T & \Sigma_{ss}^b
\end{bmatrix} \succeq 0.    
\end{align}
It was shown in Theorem 6 that $A+BK=(X_1-W_0)G_k$. Using this fact and defining $G_K=V {\Sigma_{ss}^b}^{-1}$ result in \eqref{riskda2}. Moreover, using $K=U_0 \, G_K$ and  $G_K=V {\Sigma_{ss}^b}^{-1}$, one has $K=U_0 V {\Sigma_{ss}^b}^{-1}$. This completes the proof. \hfill   $\square$

\begin{rem}
Note that Theorem 6 requires the measurements of the noise sequences during learning. The noise measurement is not needed after a safe controller is learned. The requirement of measuring the noise signal during learning can be relaxed as follows. It was shown in Theorem 6 that $A+BK=(X_1-W_0)G_K$. Since the noise is zero mean by Assumption 1, $\mathbb{E}(A+BK)=X_1 \, G_K$. Now, define $\theta=\text{vec}(A+BK)$
and $\hat \theta=\text{vec}(X_0{G_K}-W_0 G_K)$. Then, one has
\begin{align}
   & \nonumber \text{Var}(\hat \theta)=\mathbb{E} [(\theta-\hat \theta)^T (\theta-\hat \theta)]= \\  & \nonumber \mathbb{E}[\text{vec}(W_0 G_K) (\text{vec}(W_0 G_K))^T]= 
   (G_K^T \otimes I) (I \otimes \Sigma \Sigma^T)   (G_K \otimes I)=\\
   &(G_K^T G_k \otimes  \Sigma \Sigma^T)
\end{align}
where the first equality comes from $\text{vec}(W_0)=(\bar 1 \otimes \Sigma)$. Therefore, in Theorem 6, one can ignore the noise, and, instead add a soft constraint $||G_K^T G_k|| \le \gamma$ and optimize over $\gamma$ to learn a controller that achieves safety with maximum probability. The optimization problem, however, will be a semi-definite programming (SDP), instead of an LP. This will result in a minimum-variance certainty-equivalence solution
since the learning is performed as if the noise were zero. 
\end{rem}

\section{Simulation results} 
Consider a linear system in the form of \eqref{syst} with the state vector $x=[x_1,x_2,x_3,x_4]^T$ and dynamics
\begin{align}  \label{AB}
A=
\begin{bmatrix}
0.2 &   0.00 &  -0.1 &  0.0 \\
   -0.0 &  -0.200  &  0.500  & 0.1 \\
   -0.1 &  -0.5   & 1    & 0.0 \\
    0.1 &   0.4 &  -0.6 &  0.1
\end{bmatrix},
\,\,
B=
\begin{bmatrix}
1 & 1 & 0\\
0 & 1 & 1 \\
1 & 0 & 1 \\
1 & 1 & 2
\end{bmatrix}.
\end{align} 
Note that the system dynamics $A$ and $B$ are not used by the learning algorithm and are only used to generate data for learning in a simulation environment. Let the constraint on the system be described as $x_2 \le 0.1$. Even if rich data are available to learn a system model, the CBF-based methods for DT systems (in contrast to CT systems) only work for the case where the support of the noise is finite. Besides, even for a noise with finite support, the following problem must be solved \cite{SB7,SB8}
\begin{align}
  &  \inf_\eta \Big\{\eta+\frac{1}{\beta} \sum_{i=1}^{|W|} [F \,A \, x(t)+F \, B \, u(t)+w(t)-1-\eta]^{+} p(w_i) \Big \} \\ \nonumber 
    & \ge \alpha (F x(t)-1)
\end{align}
for some $\alpha \ge 1$, where $\beta$ is the confidence level, $F=[10,0,0,0], $ $|W|$ is the support of the noise, $w_i, \,\, i=1,..,|W|$ are the possible noise realizations and $p(w_i)$ is the probability of the occurrence of $w_i$. As can be seen, this optimization must be solved at every step and it is not clear how to design a feedback controller from it. It can, however, be leveraged to myopically intervene with a nominal controller to certify its safety. In contrast, the presented approach only requires to solve an LP optimization even for infinitely-supported noise distributions. \\
\indent We assume that the control input that is used for data generation is $U_0=[u(1),...,u(5)]$ with $u(i)=[0.5,0.3,0.2]^{T}, \, i=1,...,5$. To identify the system even for the deterministic case using the least squares to solve \vspace{-6pt} 
\begin{align} 
X_1=[B \,\,\, A] \left[ \begin{array}{l}
{U_0}\\
{X_0}
\end{array} \right], 
\end{align}
one needs $n+m=4+3=7$ independent samples, which means at least 7 control input sequence must be applied to the system to generate informative data for system identification. Therefore, existing CBF-based methods that rely on a model fail to learn a safe controller. On the other hand, hedging against the worst-case noise is impossible when the noise is Gaussian and thus robust control methods are not feasible. Even if the noise is truncated, the robust optimization can be efficiently solved only if noise can be limited to some special convex sets such as polyhedral sets \cite{example}. Otherwise, the LP optimization without uncertainty will turn into a non-convex optimization with uncertainty. Finally, it is not clear how to truncate the noise to assure feasibility and non-conservativeness.\\
\indent To perform the simulation, the noise is assumed Gaussian with covariance of $0.1 \, I$ and $\lambda=0.98$. For the case where the covariance is assumed known, Theorem 6 is used to learn the safe control input using only data. To assure stability besides safety, we impose a large bound on  state $x_1$, $x_3$ and $x_4$ (i.e, $x_i \le g, i=1,3,4$ for a large value of g) to make the safe set compact. The learned $G_K$ and its corresponding $K$ are given by
\begin{align}  \label{AB}
& G_k=
\begin{bmatrix}
3.4303  &  0.5066  &  1.4290  & -0.3559 \\
  -24.4818 &  -0.6606  & -17.7040 &  -2.3169\\
   60.3760  &       0  & 42.1047  &  9.4111\\
  -46.4359  &  3.5019 & -34.4125 &  -7.5888\\
    7.1114  & -2.5166 &   7.5828 &   0.6504
\end{bmatrix},
\\
& K=
\begin{bmatrix}
0  &  0.4157 &   -0.5  & -0.1 \\
   0   & 0.2494  & -0.3   & -0.06 \\
   0 &  0.1663  & -0.2 &  -0.04
\end{bmatrix}.
\end{align} 
That is, a safe control gain is learned using a set of data that is not rich enough to learn a system dynamics from. That is, in this case, the CBF-based methods fail to find a safe controller. The state trajectories of the closed-loop system are shown for 100 different realizations of the noise and starting from $x=[1, \,\, 0.1, \,\, 1, \,\, 1]^T$. As can be seen from Figure 2, the state $x_2$ never violates its safety condition. As can be seen from Figures 1-4, the system is stable under the data-based controller learned using the presented approach. 
\begin{figure}
\begin{center}
\includegraphics[width=\linewidth,height=2.0in]{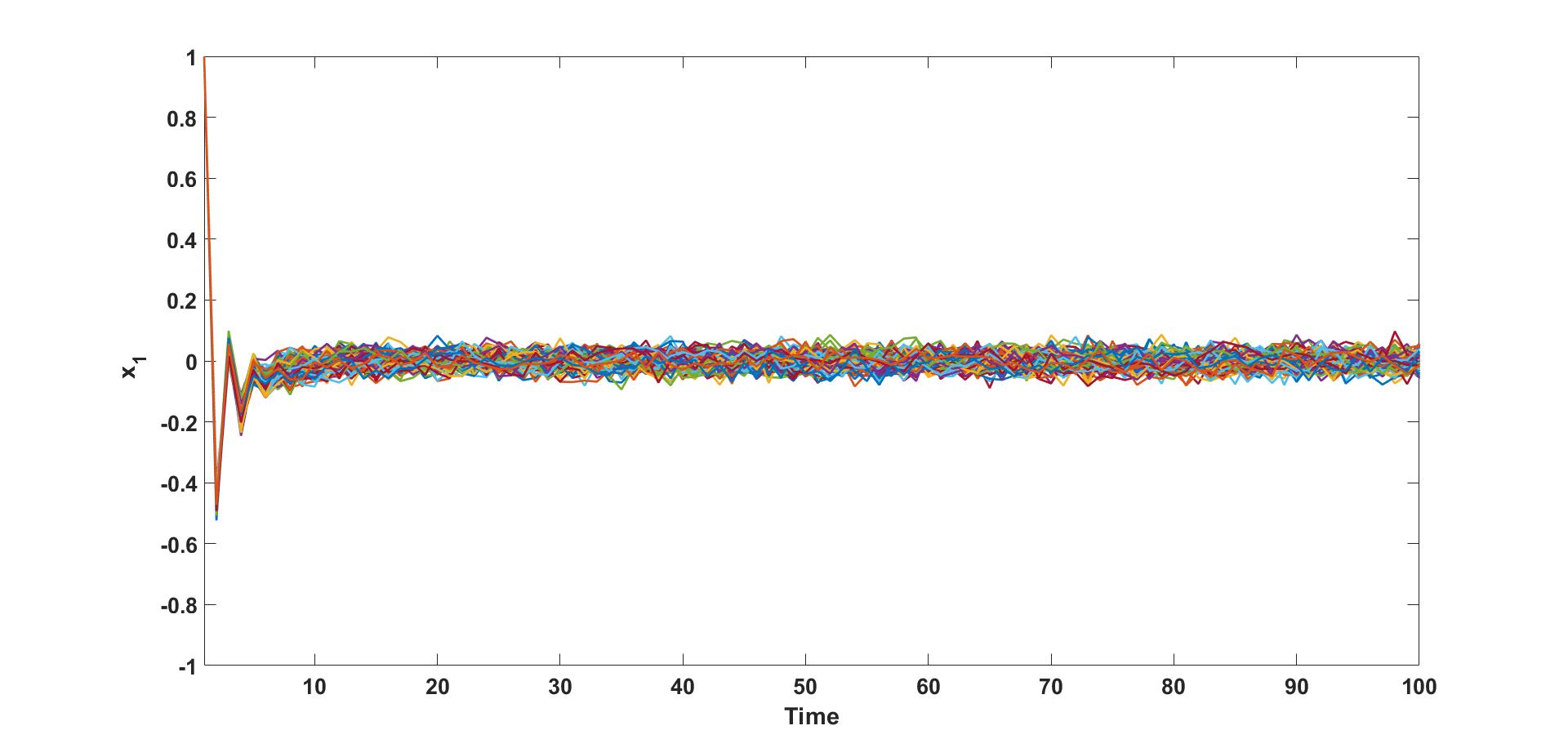}
\caption{The state trajectory $x_1$ for 100 different noise realizations}
\vspace{-10pt}
\end{center}
\end{figure}
\begin{figure}
\begin{center}
\includegraphics[width=\linewidth,height=2.0in]{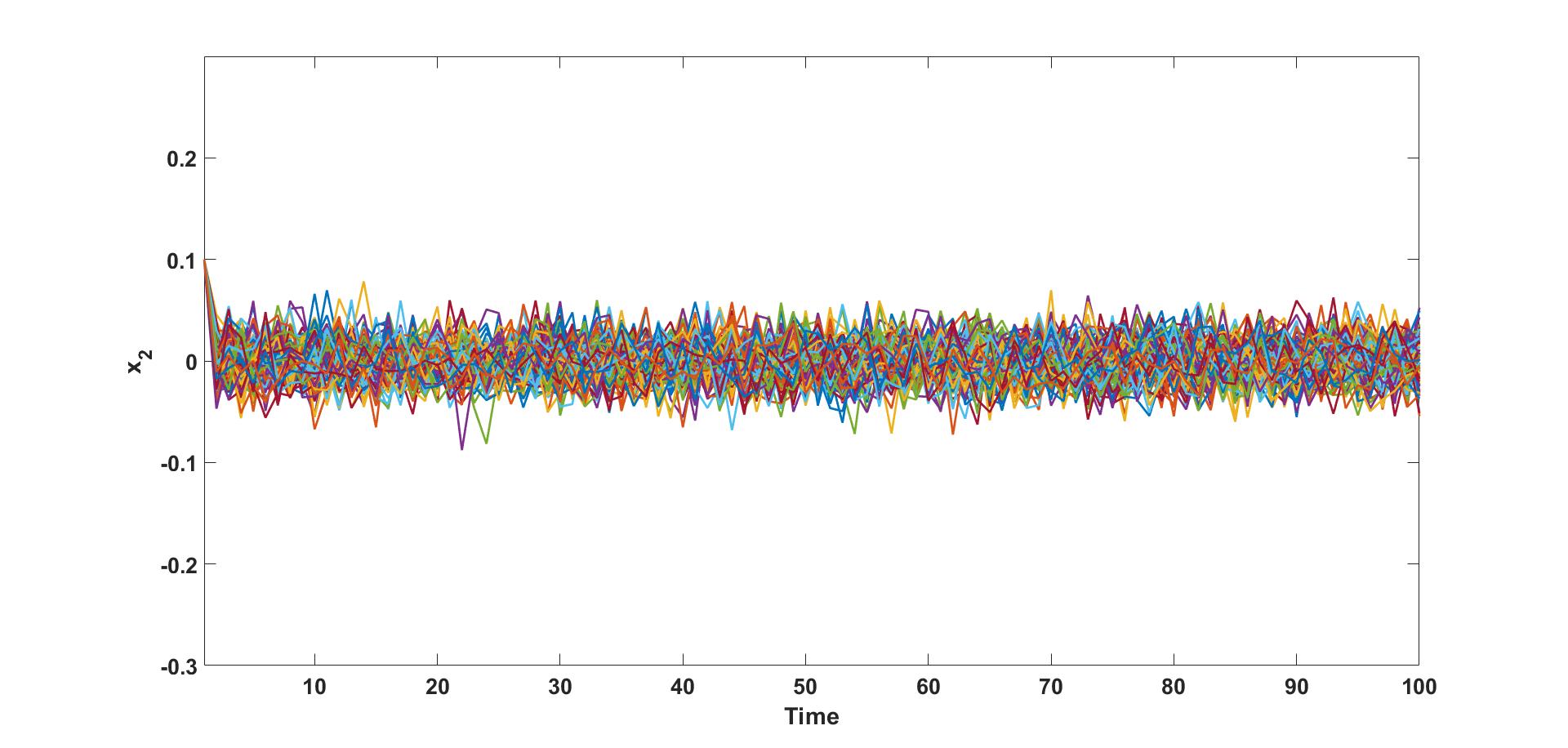}
\caption{The state trajectory $x_2$ for 100 different noise realizations}
\vspace{-10pt}
\end{center}
\end{figure}
\begin{figure}
\begin{center}
\includegraphics[width=\linewidth,height=2.0in]{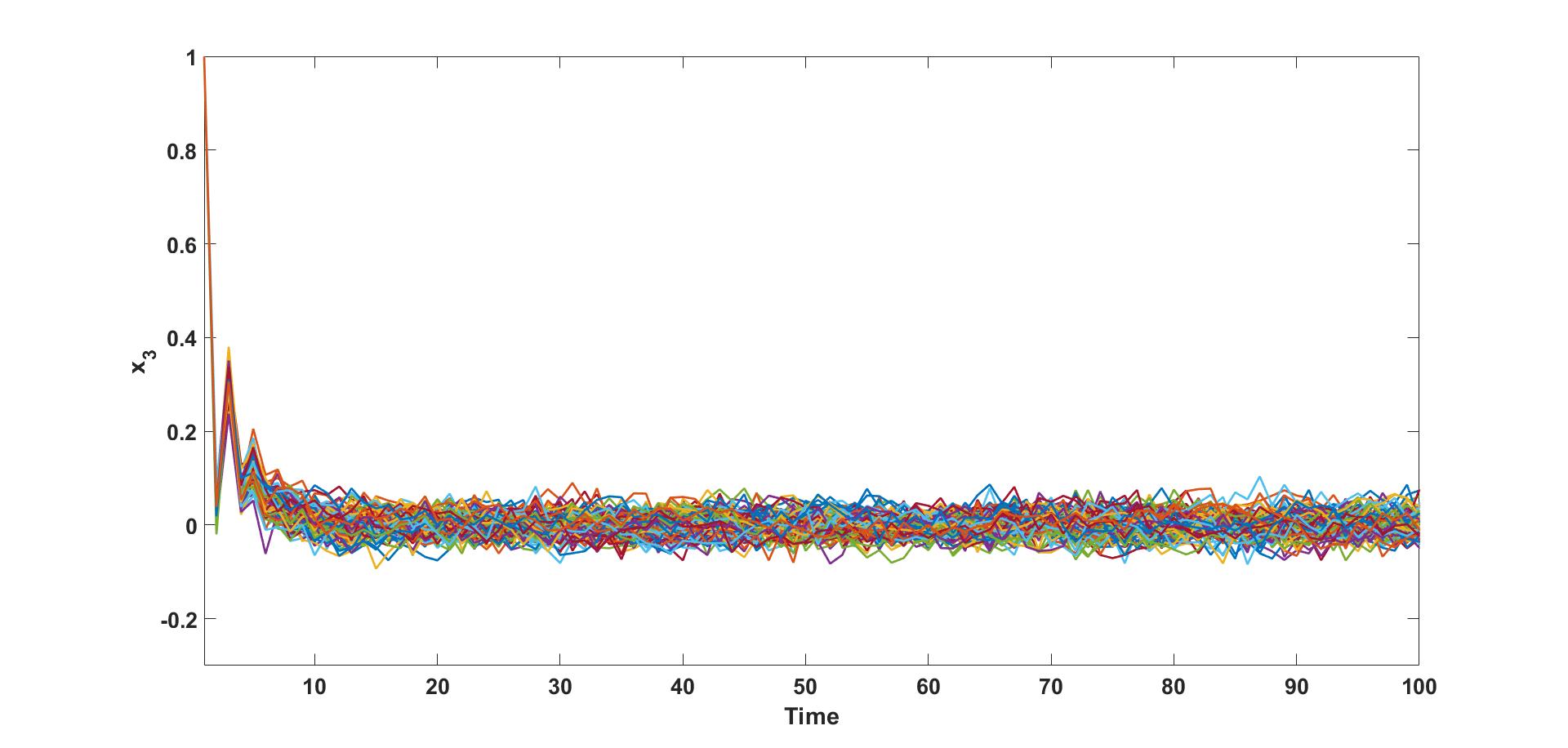}
\caption{The state trajectory $x_3$ for 100 different noise realizations}
\vspace{-1pt}
\end{center}
\end{figure}
\begin{figure}
\begin{center}
\includegraphics[width=\linewidth,height=2.0in]{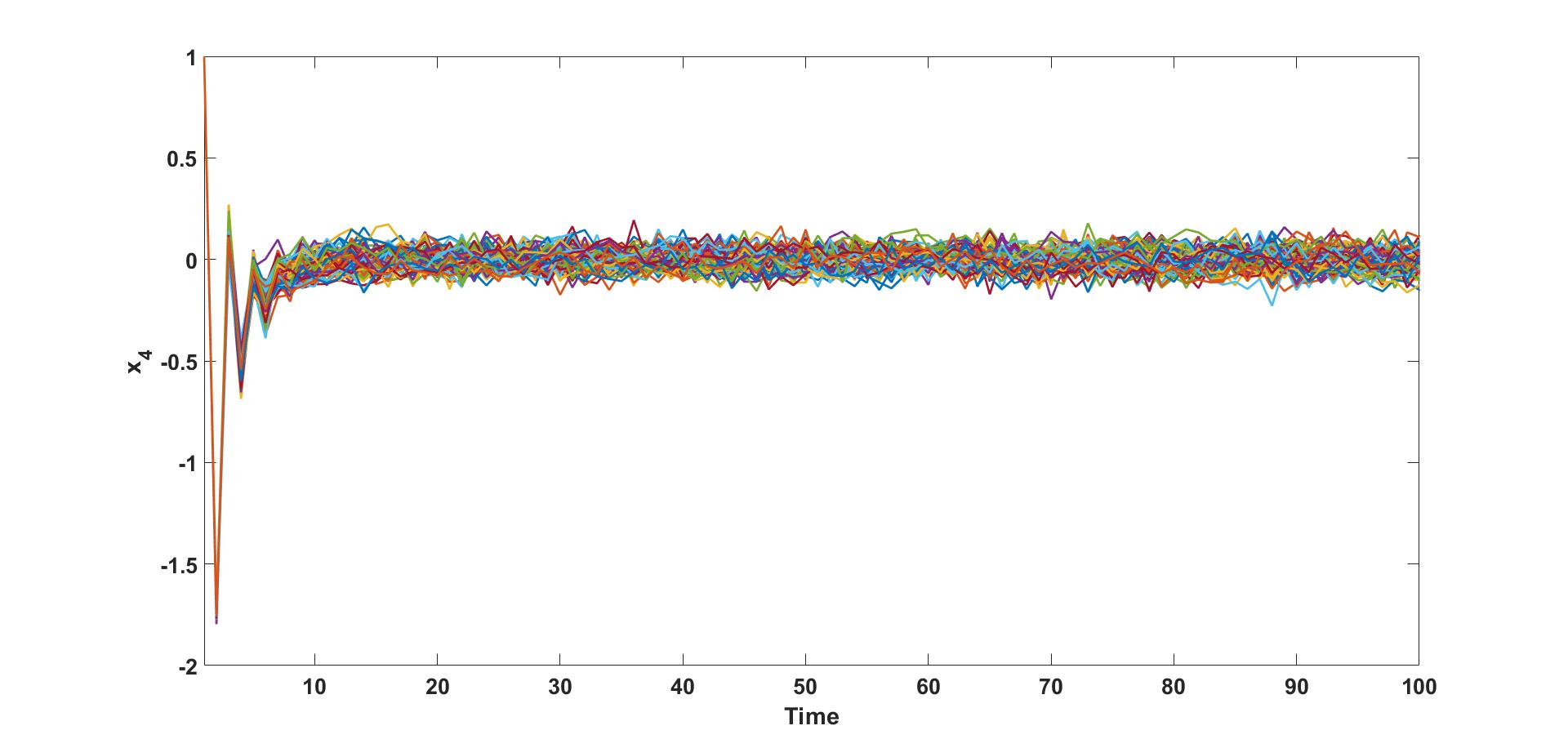}
\caption{The state trajectory $x_4$ for 100 different noise realizations}
\vspace{-20pt}
\end{center}
\end{figure}

\section{Conclusion}
Data-based safe controllers are presented for stochastic uncertain linear discrete-time systems under aleatory uncertainties. Different assumptions on the noise PDF are considered and the concept of probabilistic $\lambda$-contractive sets is leveraged to design probabilistic safe controllers using easy-to-check conditions. A bound on the risk level is first found using only the collected data and then a risk-averse safe controller is designed using only the collected data. It is also shown that directly learning a safe controller is less data-hungry and less conservative than identifying a dynamic system first and then designing a safe controller accordingly. The future work is to leverage the designed probabilistic safe controller to certify safety of reinforcement learning (RL) controllers. Rather a myopic intervention, as performed using CBF-based approaches, the learned safe controller will be merged with the RL controller.

\bibliographystyle{IEEEtran}

\end{document}